\UseRawInputEncoding
\documentclass[%
 reprint,
 amsmath,amssymb,
 aps,
]{revtex4-2}

\usepackage{graphicx}
\usepackage{dcolumn}
\usepackage{bm}
\usepackage{color}
\usepackage{epstopdf}

\begin{document}

\preprint{APS/123-QED}

\title{Full quantum theory for magnon transport in two-sublattice magnetic insulators and magnon junctions}

\author{TianYi Zhang$^1$}
\author{XiuFeng Han$^{1,2,3}$}%
 \email{xfhan@iphy.ac.cn }
\affiliation{%
 1. Beijing National Laboratory for Condensed Matter Physics, Institute of Physics, University of Chinese Academy of Sciences, Chinese Academy of Sciences, Beijing 100190, China\\
2. Center of Materials Science and Optoelectronics Engineering, University of Chinese Academy of Sciences, Beijing 100049, China\\
3. Songshan Lake Materials Laboratory, Dongguan, Guangdong 523808, China
}%



\date{\today}

\begin{abstract}
Magnon, as elementary excitation in magnetic systems, can carry and transfer angular momentum. Due to the absence of Joule heat during magnon transport, researches on magnon transport have gained considerable interests over the past decade. Recently, a full quantum theory has been employed to investigate magnon transport in ferromagnetic insulators (FMIs). However, the most commonly used magnetic insulating material in experiments, yttrium iron garnet (YIG), is a ferrimagnetic insulator (FIMI). Therefore, a full quantum theory for magnon transport in FIMI needs to be established. Here, we propose a Green's function formalism to compute the magnon bulk and interface current in both FIMIs and antiferromagnetic insulators (AFMIs). We investigate the spatial distribution and temperature dependence of magnon current in FIMIs and AFMIs generated by temperature or spin chemical potential step. In AFMIs, magnon currents generated by temperature step in the two sublattices cancel each other out. Subsequently, we numerically simulate the magnon junction effect using the Green's function formalism, and result shows near 100 \% magnon junction ratio. This study demonstrates the potential for investigating magnon transport in specific magnonic devices using a full quantum theory.
\end{abstract}

\maketitle



Magnon, which is the elementary excitation in magnetic system\cite{Chumak_2015,Yuan_2022,Bloch_1930}, has potential as information carrier due to the ability to carry and transfer angular momentum. Compared to electrons, there are three main advantages of using magnons instead to transport information: Firstly, magnon transport does not generate Joule heat. Xiao et al.'s work shows that magnons can transport in magnetic insulators \cite{Xiao_2010}, which avoids the generation of Joule heat. Secondly, magnon is an ideal carrier for transporting GHz or THz information \cite{Adam_1988,Cherepanov_1993,Owens_1985,Balashov_2014,Chuang_2014}. The coherent magnons excited in ferromagnets and antiferromagnets have ranges from GHz to THz eigenfrequency, respectively, which is the spectrum to be developed. Thirdly. There are many ways to inject and detect magnon current. The ways to inject magnons include microwave antenna \cite{Schneider_2008,Jamali_2013,Vladislav_2009}, the spin Seebeck effect (SSE) \cite{Ohe_2011,Bauer_2012,Yu_2017,Xiao_2010,Rezende_2014}, and the spin hall effect (SHE) \cite{Sinova_2015}; the detection ways include Brillouin light scattering \cite{Demokritov_2001} and the inverse spin hall effect (ISHE) \cite{Werake_2011,Saitoh_2006}. Recently, there has been an increase in studying spin transport that involves magnons, such as magnon mediate drag effect \cite{Zhang_2012,Wu_2016}, magnon valve effect \cite{Wu_2018,Cornelissen_2018,Cramer_2018}, and magnon junction effect \cite{Guo_2018}. Like metal-oxide-semiconductor field-effect transistor (MOSFET) in microelectronic devices, magnon junction, composed of a ferromagnetic insulator (FMI1)/antiferromagnetic insulator (AFMI)/ferromagnetic insulator (FMI2), is an elementary device that controls the opening and closing of magnon transport channels. To be more specific, we can control the magnitude of the output magnon current by manipulating the magnetization state of the two FMI layers. The output magnon current is larger for the parallel state and shorter for the antiparallel state. 

In order to further understand the experimental phenomenon, many theories have been proposed for studying magnon transport. For example, the LLG equation, originally introduced by Landau and Lifshitz, and later modified by Gilbert \cite{Gilbert_2004} is a classical equation widely used to calculate magnon accumulation and transport \cite{Ritzmann_2014,Ritzmann_2017}. The magnon Schrodinger equation \cite{Yan_2011,Wang_2015,Lan_2015,Jia_2019,Yu_2016,Xing_2021,Goos_2019,Lee_2017} is also used to study the wave properties of magnons and their coherent transport. The magnon Boltzmann function \cite{Manchon_2008,Cornelissen_2016,Liu_2019,Sinova_2004} is used to describe magnon transport from a particle point of view. Recently, Duine et al. propose a Green's function formalism \cite{Sterk_2021,Zheng_2017} to describe magnon transport in FMIs with and without anisotropy terms. The advantage of this approach is that the Green's function formalism provides a full quantum theory for describing magnon transport, enabling the convenient consideration of disorder and magnon coupling with other particles or quasi-particles. In experiments, one of the most commonly used magnetic materials is yttrium iron garnet (YIG), a ferrimagnetic insulator (FIMI), thus a Green's function formalism for FIMI is needed. 

In this paper, we analytically derive the Green's function formalism for magnon transport in FIMI or AFMI. We demonstrate that there are two effective magnon currents in FIMI or AFMI with no interaction between them. And it is reasonable that we could only consider on-site energy and next-nearest neighbor transition energy for these two types of magnons. We also investigate the spatial distribution of magnon current excited by temperature or spin chemical potential step in FIMI or AFMI, and we calculate the temperature dependence of magnon current, which is consistent with previous research \cite{Wu_2016}. Furthermore, we study the magnon transport in a magnon junction, simulate the magnon junction effect and the result shows near 100 \% magnon junction ratio. Our work demonstrates the possibility of using the Green's function formalism to investigate magnon transport in specific magnonic devices.


The Hamiltonian of FIMI or AFMI, considering the nearest neighbor and next-nearest neighbor Heisenberg exchange interactions, can be expressed as follows:
\begin{equation} \label{GrindEQ__1_} 
\begin{array}{rc}
\hat{H}= & -J_{A B} \sum_{< i, m>} \boldsymbol{\hat{S}}_{i} \cdot \boldsymbol{\hat{S}}_{m}-J_{A} \sum_{\ll i, j \gg} \boldsymbol{\hat{S}}_{\boldsymbol{i}} \cdot \boldsymbol{\hat{S}}_{j} \\
& -J_{B} \sum_{\ll m, n \gg} \boldsymbol{\hat{S}}_{m} \cdot \boldsymbol{\hat{S}}_{\boldsymbol{n}}-h_{\text {ext }}(\sum_{i} \mu_{A} \hat{S}_{i}^{z} \\
& +\sum_{m} \mu_{B} \hat{S}_{m}^{z})
\end{array}
\end{equation}
Where ${<}$ ${>}$ denotes summing over nearest sites, ${\ll}$ ${\gg}$ denotes summing over next-nearest sites. $J_{AB}$${}_{\ }$and $J_{A\left(B\right)}$ represent the nearest and next-nearest Heisenberg exchange interactions strength, respectively. ${\boldsymbol{S}}_{\boldsymbol{{i}}\left(\boldsymbol{{m}}\right)}$ is the spin in A(B) sublattice, ${\mu }_{A\left(B\right)}$ is the magnetic moment in A(B) sublattice. $h_{ext}$${}_{\ }$is applied magnetic field along the z direction. Using Holstein-Primakoff (HP) transformation \cite{Yuan_2022}, Fourier transformation and Bogoliubov transformation (Details in Supplemental Material \cite{sup}) we can get
\begin{equation} \label{GrindEQ__2_} 
 \begin{array}{ll}
\hat{H} &{=}{\mathit{\sum}}_k[\left[\frac{A_k{-}B_k}{{2}}{+}\frac{\sqrt{\left(A_k{+}B_k\right)^2{-}{4}C^{{2}}_k}}{{2}}\right]\hat{\alpha }^{{\dagger }}_k\hat{\alpha }_k \\
 &\quad {+}\left[\frac{{-}A_k{+}B_k}{{2}}{+}\frac{\sqrt{\left(A_k{+}B_k\right)^2{-}{4}C^{{2}}_k}}{{2}}\right]\hat{\beta }^{{\dagger }}_k\hat{\beta }_k]{+}const \\ 
 & {\equiv }\sum_k({w^\alpha _k\hat{\alpha }^{{\dagger }}_k\hat{\alpha }_k{+}w^\beta_k\hat{\beta }^{{\dagger }}_k\hat{\beta }_k})+const \end{array}
\end{equation}
where $\hat{\alpha}_k\left(\hat{\beta}_k\right),\hat{\alpha}^{\dagger }_k\left(\hat{\beta}^{\dagger }_k\right)$ are magnon annihilation and creation operators in A(B) sublattice, respectively. $A_k\equiv -2J_AS_A{\gamma }_{k,nn}-J_{AB}S_BN_n+2J_AS_AN_{nn}+h_{ext}{\mu }_A,\ B_k\equiv -2J_BS_B{\gamma }_{k,nn}-J_{AB}S_AN_n+2J_BS_BN_{nn}-h_{ext}{\mu }_B,\ C_k\equiv -J_{AB}\sqrt{S_AS_B}{\gamma }_{k,n}$, $\ N_n$, $N_{nn}$ are the numbers of nearest and the next-nearest sites, respectively. In the case of one-dimensional atomic chain model, $N_n=N_{nn}=2$, ${\gamma }_{k,n}=2cos\left(ka\right)$, ${\gamma }_{k,nn}=2cos\left(2ka\right)$, where $a$ is the distance between nearest sites.
Thus, in both FIMI or AFMI, the magnon currents can be separated into two uncoupled magnon currents with opposite polarity. Specifically, in AFMI, $J_A=J_B,S_A=S_B,{\mu }_A={\mu }_B$, but in FIMI, the above equation does not hold.

Eq. (2) shows that in FIMI or AFMI, considering the nearest neighbor and next-nearest neighbor Heisenberg exchange interactions, it can still be thought there are two independent types of magnons in it. We can use Fourier transformation to Eq. (2) to transform Hamiltonian of FIMI or AFMI to coordinate space, and get
\begin{equation} \label{GrindEQ__3_} 
w^{\alpha \left(\beta \right)}_k{=}\sum^{{\infty }}_{n{=0}}{{2}A_n\left(B_n\right){cos \left(nka\right)\ }} 
\end{equation}
Where $A_0\ \left(B_0\right)$ and $A_1\ \left(B_1\right)$ represent on-site and nearest transition energy of magnons in A (B) sublattice. Then the Hamiltonian can be written as
\begin{equation} \label{GrindEQ__4_} 
\hat{H}{=}\sum^{{\infty }}_{n{=0}}{\sum_{i,j}{\delta \left(i{-}j{\pm }n\right)\left(A_n{\alpha }^{{\dagger }}_i{\alpha }_j{+}B_n{\beta }^{{\dagger }}_i{\beta }_j\right)}} 
\end{equation}
From Eq. (2${\sim}$4), we can see that for one-dimensional atomic chain model $A_{2n+1}=B_{2n+1}=0$(n = 0, 1, 2 $\cdots$), which means that magnons can only propagate between sites that are an even number of lattice constants away from each other. We investigate the variation of the Fourier expansion coefficient with expansion order, where for AFMI we use the following parameters \cite{Kodderitzsch_2002,Zhang_2006}: $J_{AB} = -0.002$ eV, $J_A = J_B = 0.02$ eV, $S_A=S_B=1$, and for FIMI we use the following plausible parameters \cite{Ritzmann_2017}: $J_{AB} = -0.005$ eV, $J_A = 0.05$ eV, $J_B = 0.01$ eV,$S_A=1$, $S_B=1.5$ , as shown in Fig. 1. We can see that for both AFMI and FIMI the odd parts of the expansion coefficients are consistently 0, while the even parts approach 0 when n ${>}$ 4, indicating that only two terms $A_0\ \left(B_0\right)$ and $A_2\ \left(B_2\right)$ need to be retained. 
\begin{figure}[htbp]
\includegraphics[width = \linewidth]{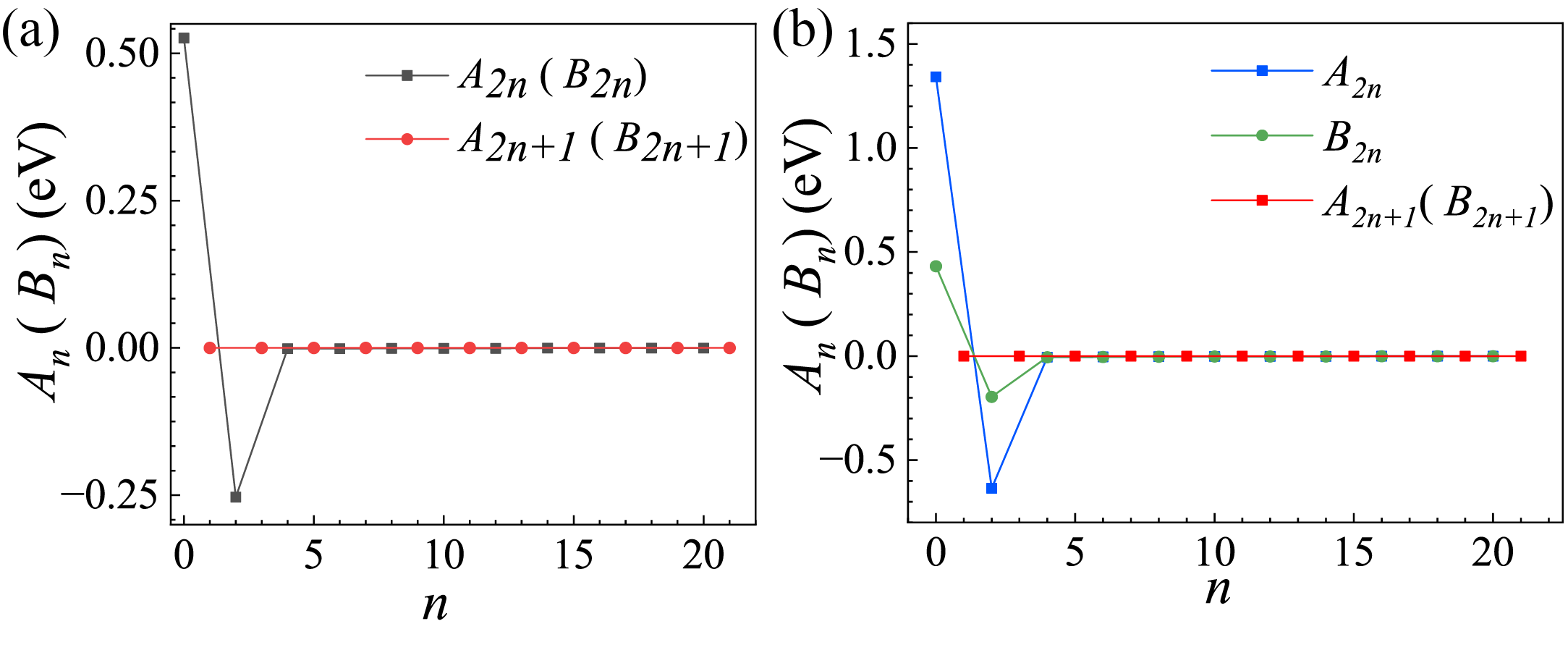}
\caption{ The variation of the Fourier expansion coefficient with expansion order {n} for AFMI (a) and FIMI (b).}
\end{figure}

Then we study the magnon transport in FIMI or AFMI, only two terms $A_0\ \left(B_0\right)$ and $A_2\ \left(B_2\right)$ are taken into account. A schematic diagram that illustrates the transport of magnon current through FIMI or AFMI is shown in Fig. 2. In this setup, the FIMI or AFMI is connected to two heavy metals (HMs) with temperatures {T${}_{R}$}, {T${}_{L}$} and spin chemical potentials {$\mu$${}_{L}$}, {$\mu$${}_{R}$}, respectively. The magnon current is driven by the difference of temperature or spin chemical potential between two HMs. To calculate the bulk magnon current, we need to calculate the retarded and advanced Green's functions firstly. This can be accomplished by using the Dyson equation
\begin{equation} \label{GrindEQ__5_}
{\mathcal{G}}^{R\left(A\right)}{(}\varepsilon {)=}{\left[\varepsilon^{+(-)} {-}H{-}{\hslash }{\mathit{\Sigma}}^{R\left(A\right)}\left(\varepsilon \right)\right]}^{{-}{1}}
\end{equation}
Where ${\mathcal{G}}^{R\left(A\right)}$ is retarded (advanced) Green's function, $\varepsilon^{+(-)}=\varepsilon+(-)i\eta$, $\eta$ is a infinitesimal positive number, {H} is Hamiltonian of magnons in FIMI or AFMI, and ${\mathit{\Sigma}}^{R\left(A\right)}\left(\varepsilon \right)$ is retarded (advanced) self-energy, which describes the coupling between FIMI or AFMI and external system.
\begin{figure}[htbp]
\includegraphics[width = \linewidth]{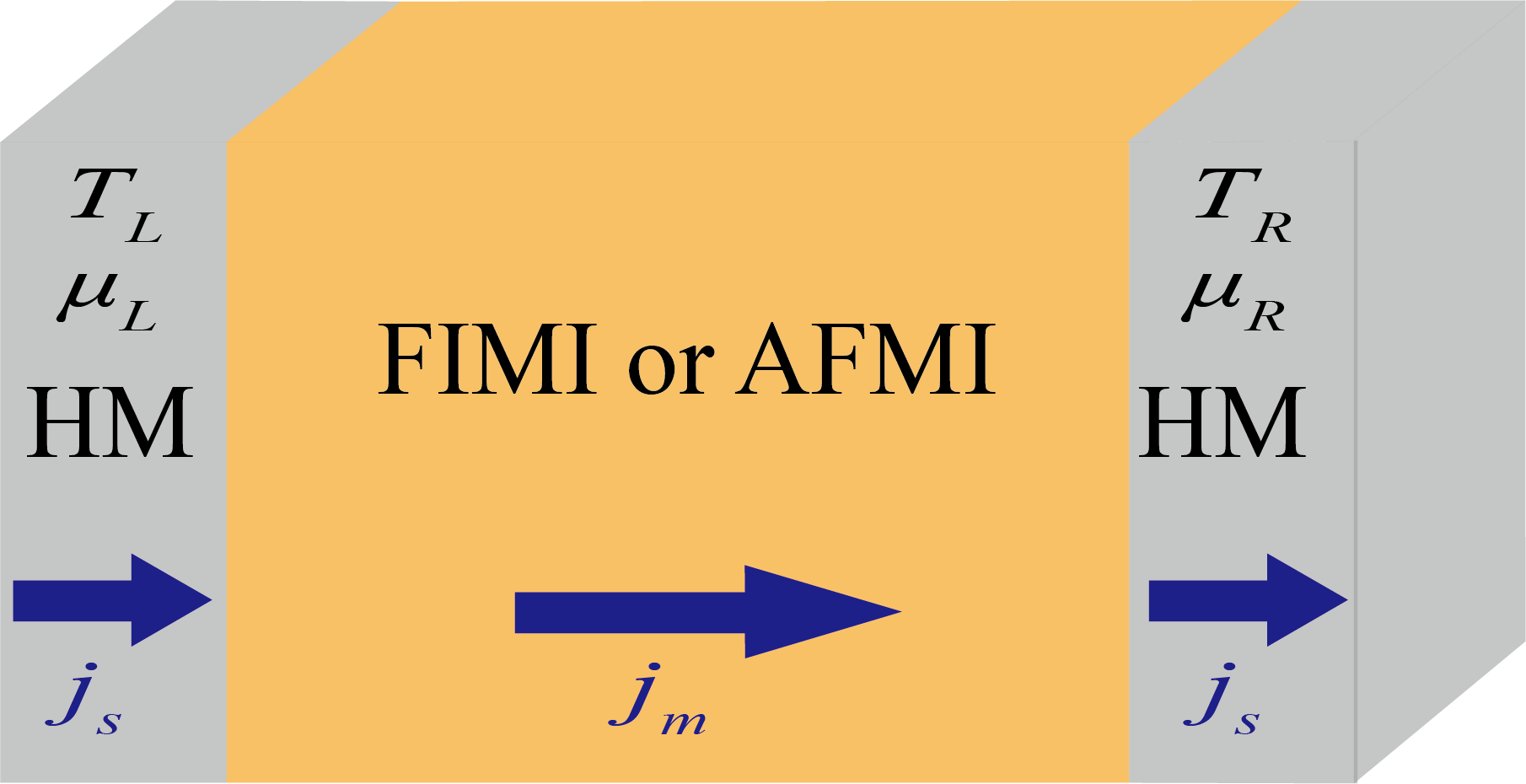}
\caption{ Schematic diagram that illustrates the transport of magnon current through FIMI or AFMI driven by temperature or spin chemical potential difference.}
\end{figure}
For FIMI or AFMI, the Hamiltonian is
\begin{equation} \label{GrindEQ__6_} 
\begin{array}{ll}
\hat{H}=&\sum_{i,j}[{(A_0\ {\delta }_{i,j}+A_2\ {\delta }_{i,j\pm 2})\hat{\alpha }^{{\dagger }}_i\hat{\alpha }_j}\\
&+{(B_0{\delta }_{i,j}+B_2{\delta }_{i,j\pm 2})\hat{\beta }^{{\dagger }}_i\hat{\beta }_j}]
\end{array}
\end{equation}
and the self-energy are composed of three items ${{\Sigma }}^{R\left(A\right)}\left({\varepsilon }\right)={{\Sigma }}^{R\left(A\right)}_C\left({\varepsilon }\right)+\sum_{r{\in }\left\{L,R\right\}}{{{\Sigma }}^{R\left(A\right)}_r\left({\varepsilon }\right)}$, where
\[{{\Sigma }}^{{R}\left({A}\right)}_{{C\ i,j}}\left(\varepsilon \right){=}{\mp }i\alpha (\varepsilon-\mu_C) {\delta }_{i,j}{/}{\hbar },\] 
\[{\mathit{\Sigma}}^{R\left(A\right)}_{L{\ }{\ }i,j}{=}{\mp }i{\eta }^L\left(\varepsilon {-}{\mu }_L\right){\delta }_{i,j}\left({\delta }_{j{,1}}{+}{\delta }_{j{,2}}\right){/}{\hbar }\] 
\begin{equation} \label{GrindEQ__7_} 
{\mathit{\Sigma}}^{R\left(A\right)}_{R{\ }{\ }i,j}{=}{\mp }i{\eta }^R\left(\varepsilon {-}{\mu }_R\right){\delta }_{i,j}\left({\delta }_{j,N}{+}{\delta }_{j,N{-}{1}}\right) {/}{\hbar }
\end{equation} 
are retarded (advanced) self-energy induced by Gilbert damping in FIMI or AFMI, connection with left HM and right HM, respectively. Where $\alpha $ is Gilbert damping in FIMI or AFMI, ${\hslash }$ is reduced Planck's constant, ${\eta }^{L(R)}$ is parameter that shows the coupling with left and right HMs \cite{Zheng_2017,Tserkovnyak_2002}, ${\mu }_{L(R)}$ is spin chemical potential of left(right) HM, $\mu_C$ is magnon potential of FIMI or AFMI.

Secondly, we can calculate the magnon density matrix using langreth rule \cite{Zheng_2017}
\begin{equation} \label{GrindEQ__8_} 
\rho {=}\int^{{\infty }}_{{-}{\infty }}{\frac{d\varepsilon }{{2}\pi }\left[{\mathcal{G}}^R\left(\varepsilon \right)i{\hslash }{\mathit{\Sigma}}^{{<}}\left(\varepsilon \right){\mathcal{G}}^A\left(\varepsilon \right)\right]} 
\end{equation} 
where the less self-energy can be calculated by
\begin{equation} \label{GrindEQ__9_}
 \begin{array}{ll}
{\mathit{\Sigma}}^{{<}}\left(\varepsilon \right)&{=}{\mathit{\Sigma}}^{{<}}_{C}\left(\varepsilon \right){+}\sum_{r{\in }\left\{L,R\right\}}{{\mathit{\Sigma}}^{{<}}_r\left(\varepsilon \right)}\\
&{=2}iN_B\left(\frac{\varepsilon-\mu_C  }{k_BT_{C}}\right)Im\left({\mathit{\Sigma}}^R_{C}\left(\varepsilon \right)\right)\\
&{+}\sum_{r{\in }\left\{L,R\right\}}{{2}iN_B\left(\frac{\varepsilon {-}{\mu }_r}{k_BT_r}\right)Im\left({\mathit{\Sigma}}^R_r\left(\varepsilon \right)\right)}
\end{array}
\end{equation}
Where $N_B\left(x\right)=\frac{1}{e^x-1}$ is Bose-Einstein distribution. Then we can calculate bulk magnon current using Heisenberg motion equation.
\begin{equation} \label{GrindEQ__10_} 
{\hslash }\frac{d<{\alpha }^{{\dagger }}_i{\alpha }_i>}{dt}=\sum_j{-i\left(h_{i,j}{\rho }_{j,i}-h_{j,i}{\rho }_{i,j}\right)}=\sum_j{j_{m;i,j}} 
\end{equation}
where $j_{m;i,j}$ represents the magnon current from site i to site j, $h$ is Hamiltonian for $\alpha$ mode magnons. Eq. (10) indicates that the change in the magnon number at site i is caused by all magnon currents from site i to other sites.

According to \cite{Wu_2016}, the magnetization reversal of the ferromagnetic (FM) layer has no effect on the transport of magnon current induced by the spin Hall effect (SHE). However, for magnon current induced by the spin Seebeck effect (SSE), the opposite magnetization of the FM generates magnons of different signs, leading to a output voltage signal with opposite signs. Therefore, we assume that magnons with opposite polarity experience an equivalent spin voltage but an opposite temperature gradient.

The Green's function formalism can also be utilized to calculate the interface magnon current. By using the Landauer-B{\"u}ttiker formula \cite{Zheng_2017}, we find that the magnon current at the interface between FIMI or AFMI and HMs can be expressed as follows:
\begin{equation} \label{GrindEQ__11_}
\begin{array}{ll}
{{j}}^{{m}}_{{L(R)}}&{=}{{j}}^{{m}}_{{L(R),}\alpha}{+}{{j}}^{{m}}_{{L(R),}\beta}\\
&{=}\int{\frac{d\varepsilon }{{2}\pi }}\left[N_B\left({\ }\frac{\varepsilon {-}{\mu }_{L(R)}}{k_BT_{L(R)}}\right){-}N_B\left({\ }\frac{\varepsilon {-}{\mu }_R(L)}{k_BT_R(L)}\right)\right]T_{b,\alpha }\left(\varepsilon \right)\\
&{+\ }\int{\frac{d\varepsilon }{{2}\pi }}\left[N_B\left({\ }\frac{\varepsilon {-}{\mu }_{L(R)}}{k_BT_{L(R)}}\right){-}N_B\left({\ }\frac{\varepsilon {-}{\mu }_C}{k_BT_{AFMI}}\right)\right]T_{f,\alpha }\left(\varepsilon \right)\\&{+}\int{\frac{d\varepsilon }{{2}\pi }}\left[N_B\left({\ }\frac{\varepsilon {-}{\mu }_{L(R)}}{k_BT_R(L)}\right){-}N_B\left({\ }\frac{\varepsilon {-}{\mu }_R(L)}{k_BT_{L(R)}}\right)\right]T_{b,\beta }\left(\varepsilon \right)\\&{+}\int{\frac{d\varepsilon }{{2}\pi }}\left[N_B\left({\ }\frac{\varepsilon {-}{\mu }_{L(R)}}{k_BT_{AFMI}}\right){-}N_B\left({\ }\frac{\varepsilon {-}{\mu }_C}{k_BT_{L(R)}}\right)\right]T_{f,\beta }\left(\varepsilon \right)\}
\end{array}
\end{equation}
Where the transmission function
\begin{equation} \label{GrindEQ__12_}
\begin{array}{ll}
T_{b,i}\left(\varepsilon \right)&{\equiv }Tr\left[{\hslash }{\mathit{\Gamma}}_{L(R),i}\left(\varepsilon \right){\mathcal{G}}^R_i\left(\varepsilon \right){\hslash }{\mathit{\Gamma}}_{R(L),i}\left(\varepsilon \right){\mathcal{G}}^A_i\left(\varepsilon \right)\right],\\
T_{f,i}\left(\varepsilon \right)&{\equiv }Tr\left[{\hslash }{\mathit{\Gamma}}_{L(R),i}\left(\varepsilon \right){\mathcal{G}}^R_i\left(\varepsilon \right){\hslash }{\mathit{\Gamma}}_{AFMI}\left(\varepsilon \right){\mathcal{G}}^A_i\left(\varepsilon \right)\right]   
\end{array}
\end{equation}
Where $i=\alpha \ or\ \beta $, are two modes of magnons with opposite polarity, the rates ${\mathit{\Gamma}}_{L(R),i}\left(\varepsilon \right)=-2Im \left({\mathit{\Sigma}}^R_{L(R),i}\left(\varepsilon \right)\right)$.

The Green's function formalism can be utilized to calculate magnon current driven by different driving mechanisms, such as temperature difference or spin chemical potential difference. In particular, we use temperature difference between left and right HMs to simulate magnon current excited by SEE, and use spin chemical difference between left and right HMs to simulate magnon current excited by SHE. And we use a one-dimensional atomic chain model for simplicity.


To investigate the spatial distribution and temperature dependence of magnon currents in AFMI excited by the SSE and the SHE, we set the temperature difference ($\mathit{\Delta}T$) and spin chemical potential difference ($\mathit{\Delta}$$\mu$) between two HMs. The corresponding results are shown in Fig. 3.
\begin{figure}[htbp]
\includegraphics[width = \linewidth]{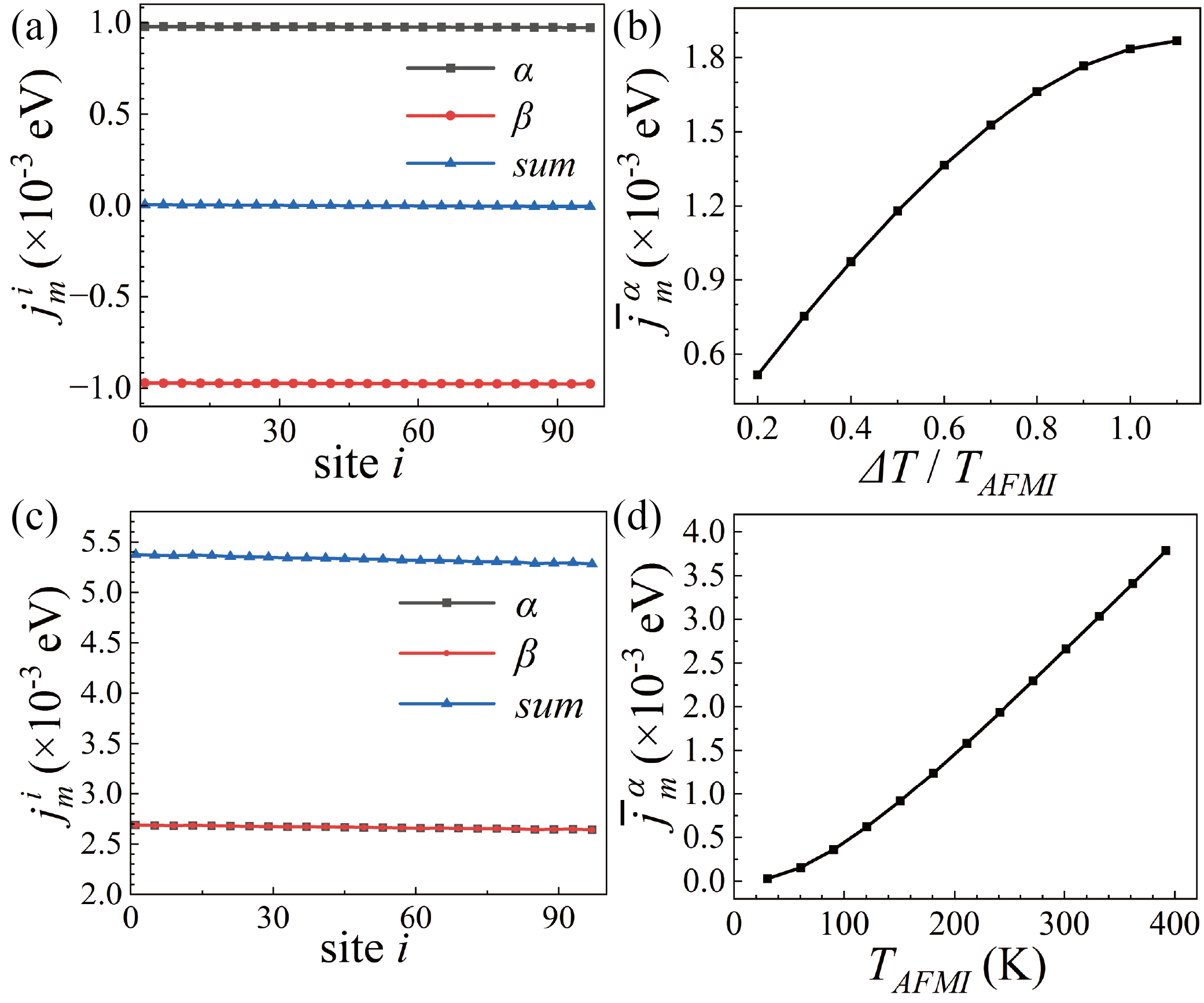}
\caption{Spatial distribution and temperature dependence of magnon currents excited by SSE (a, b) and SHE (c, d) in AFMI.}
\end{figure}
In Fig. 3. (a), we calculated spatial distribution of magnon currents excited by SSE in AFMI. The parameters used in simulation are set to be $A_0 = B_0 = 0.5$ eV, $A_2 = B_2 = -0.25$ eV, total site number $N$ = 100, applied magnetic field $h_{ext}$ = 0, spin chemical potential ${\mu }_L$ = ${\mu }_R = 0$, ${\eta }^L = {\eta }^R = 8$, temperature $k_BT_{AFMI} = 0.26$ eV, $T_L = 1.2{\ T}_{AFMI}$, $T_R = 0.8{\ T}_{AFMI}$, magnon potential $\mu_{AFMI} = 0$ and Gilbert damping ${\alpha }_{AFMI} = 0.001$ \cite{Moriyama_2019}. We can see that the magnon currents composed of {$\alpha$} and {$\beta$} modes magnons have different sign but the same absolute value, so the $\alpha$ and $\beta$ mode magnon currents cancel with each other, total magnon currrent $j^{sum}_m$ is 0. Then we keep the temperature of left HM $T_L = 1.2{\ T}_{AFMI}$ fixed, change the temperature of right HM $T_R$, and average all the $\alpha$ mode magnon current of 100 sites, the temperature dependence of averaged $\alpha$ mode magnon currents ${\overline{j}}^{\alpha }_m$ is shown in Fig. 3. (b), we can see that ${\overline{j}}^{\alpha }_m$ shows positive correlation dependence on temperature difference between left and right HMs $\Delta T = T_L-T_R$ and the influence of $\Delta T$ on ${\overline{j}}^{\alpha }_m$ is gradually reduced as $\Delta T$ increases. Then we calculated spatial distribution of magnon currents excited by SHE in AFMI, see Fig. 3. (c). Spin chemical potential ${\mu }_L=0.1 A_0$, ${\mu }_R=0$, temperature $k_BT_{AFMI}=0.026$ eV, $T_L=T_R=T_{AFMI}$. We can see that for magnon current excited by SHE, $\alpha$ and $\beta$ mode magnons contribute equally in the component of sum magnon currents. And then we change the temperature of left HM, AFMI and right HM at the same time, and calculate the temperature dependence of average $\alpha$ mode magnon current, as shown in Fig. 3. (d). We can see that  ${\overline{j}}^{\alpha }_m$ increases as $T_{AFMI}$ increases. It can be explained by that as $T_{AFMI}$ increase, the number of $\alpha$ mode magnon in sublattice $n_{\alpha}(\varepsilon) = \frac{1}{e^{\frac{\varepsilon}{k_BT_{AFMI}}}-1}$ increases, therefore, the magnons involved in transport increase.

Then we calculate the spatial distribution and temperature dependence of magnon currents in FIMI excited by the SSE and the SHE, as shown in Fig. 4. In Fig. 4. (a), we calculated spatial distribution of magnon currents excited by SSE in FIMI. The parameters used in simulation are set to be  $A_0 = 1.3$ eV, $B_0 = 0.43$ eV, $A_2 = -0.6$ eV, $B_2 = -0.2$ eV, site number {N} = 100, applied magnetic field $h_{ext}$ = 0, spin chemical potential ${\mu }_L$ = ${\mu }_R = 0$, ${\eta }^L = {\eta }^R = 8$, temperature $k_BT_{AFMI} = 0.026$ eV, $T_L = 1.2{\ T}_{AFMI}$, $T_R = 0.8{\ T}_{AFMI}$, magnon potential $\mu_{FIMI} = 0$ and Gilbert damping ${\alpha }_{FIMI} = 0.001$. We can see that although magnon currents generated by $\alpha$ and $\beta$ mode magnons have opposite sign, they do not cancel with each other, which means sum magnon current $j^{sum}_m$ is not equal to 0 in FIMI. Then we keep left HM temperature $T_L=1.2{\ T}_{AFMI}$ fixed, change the right HM temperature $T_R$, calculate the site average magnon current ${\overline{j}}^{\alpha }_m,\ \ {\overline{j}}^{\beta }_m,\ {\overline{j}}^{sum}_m$ dependence on the temperature difference, see Fig. 4. (b) we can see the absolute value of ${\overline{j}}^{\alpha }_m,\ \ {\overline{j}}^{\beta }_m,\ {\overline{j}}^{sum}_m$ increase as temperature difference between left and right HMs $\Delta T$ increases. As for magnon current excited by SHE in FIMI, we set parameters to be spin chemical potential ${\mu }_L=0.1 A_0$, ${\mu }_R=0$, temperature $k_BT_{AFMI}=0.026$ eV, $T_L=T_R=T_{AFMI}$, and calculate the space distribution of magnon current. We can see from Fig. 4. (c) that in FIMI due to the difference of on-site and next-nearest transition energy between $\alpha$ and $\beta$ mode magnons, the magnon currents composed by these two types of magnons are not the same. And then we change the temperature of the whole system at the same time, and calculate the temperature dependence of average magnon current ${\overline{j}}^{\alpha }_m,\ \ {\overline{j}}^{\beta }_m,\ {\overline{j}}^{sum}_m$. We can see from Fig. 4. (d) that ${\overline{j}}^{\alpha }_m,\ \ {\overline{j}}^{\beta }_m,\ {\overline{j}}^{sum}_m$ all increase as system temperature increase, which is due to the increase of magnons in two sublattices.
\begin{figure}[htbp]
\includegraphics[width = \linewidth]{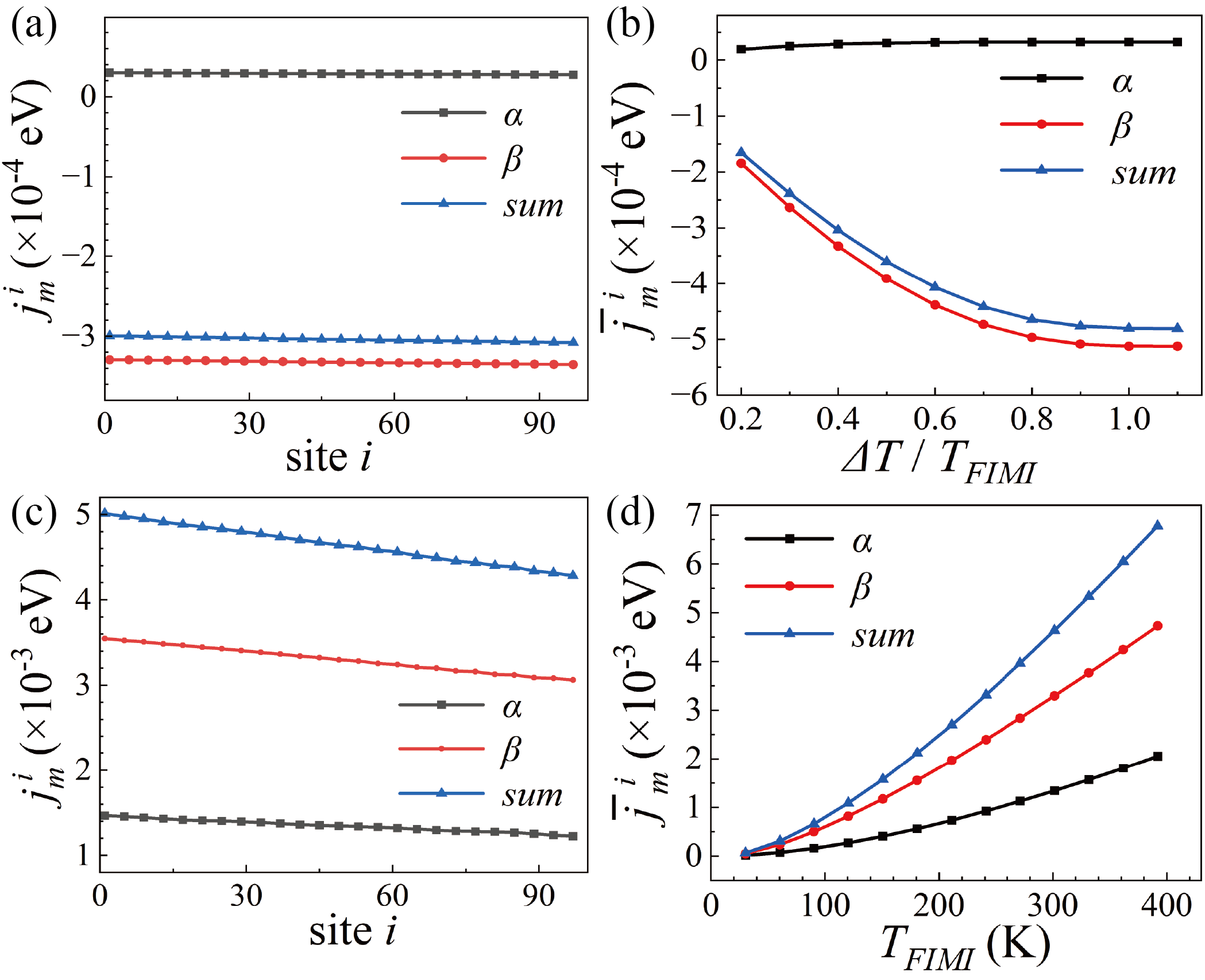}
\caption{ Spatial distribution and temperature dependence of magnon currents excited by SSE (a, b) and SHE (c, d) in FIMI.}
\end{figure}

Magnon junction is a magnon-conductive device that consists of an FMI1/AFMI/FMI2 structure with a high on-off ratio between parallel and antiparallel state of two FMI layers. We can use Green's function formalism to simulate magnon junction effect. The model includes a magnon junction and two HM leads, as shown in Fig. 5. By fixing the magnetization of FMI1 in the upward direction and varying the magnetization of FMI2 between up and down state, we can set the magnon junction to be parallel or antiparallel state.
\begin{figure}[htbp]
\includegraphics[width = \linewidth]{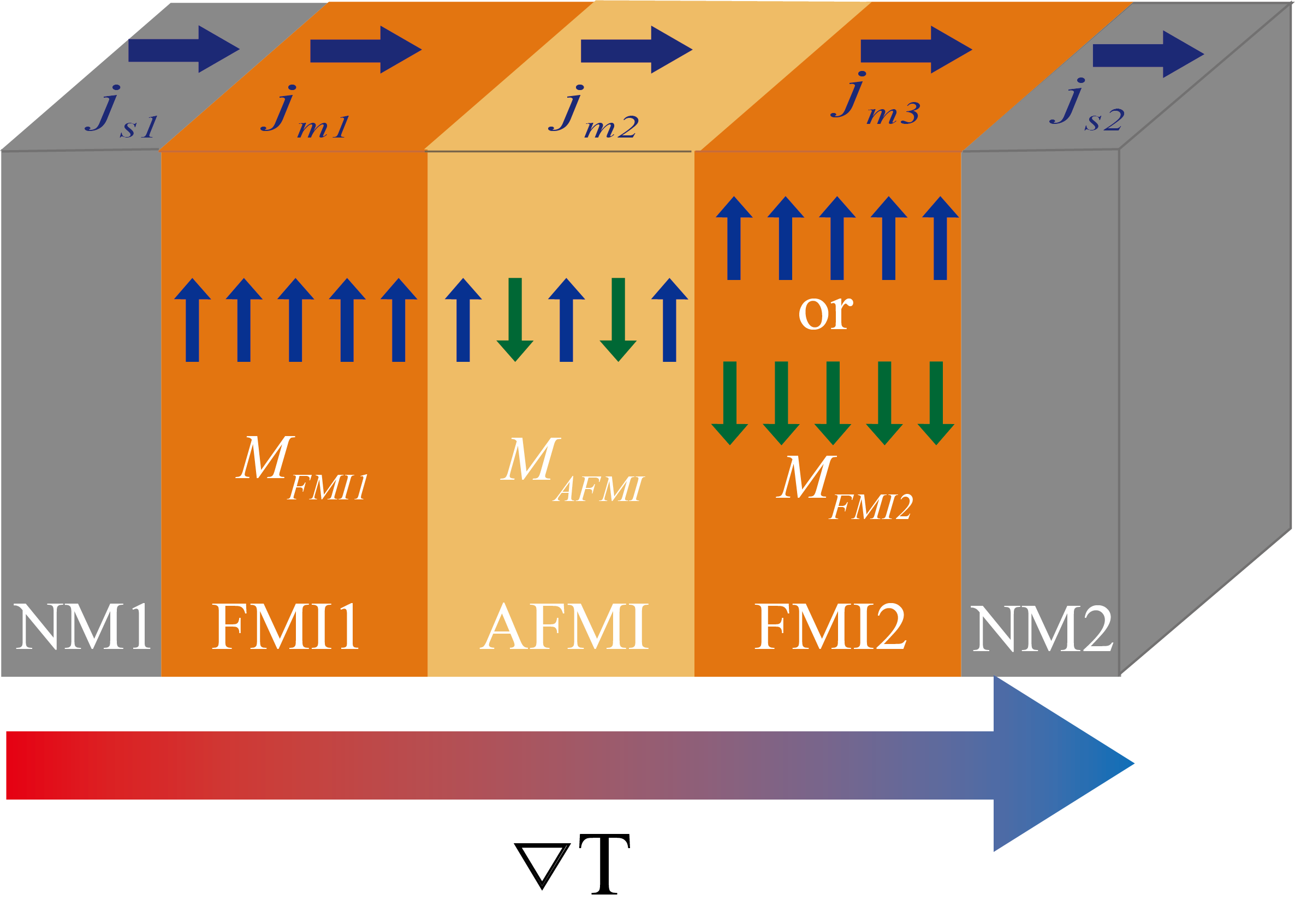}
\caption{ Schematic diagram of magnon current driven by temperature gradient transporting through a magnon junction.}
\end{figure}
The Hamiltonian of the magnon junction is composed of five items
\begin{equation} \label{GrindEQ__13_}
\begin{array}{ll}
\hat{H}=&\hat{H}_{FMI1}+\hat{H}_{AFMI}+\hat{H}_{FMI2}\\
&+\hat{H}_{FMI1,AFMI}+\hat{H}_{FMI2,AFMI}
\end{array}
\end{equation}
Where $\hat{H}_{FMI1}$, $\hat{H}_{AFMI}$, $\hat{H}_{FMI2}$ are Hamiltonian of FMI1, AFMI and FMI2, respectively, and $\hat{H}_{FMI1,AFMI}$, $\hat{H}_{FMI2,AFMI}$ are coupling between FMI1 and AFMI, FMI2 and AFMI, respectively. Only on-site and next-nearest transition energy are considered. For parallel state:
\begin{equation} \label{GrindEQ__14_} 
\hat{H}_{FMI1(2)}=\sum_{i,j}[(A^{FMI1(2)}_0\ {\delta }_{i,j}+A^{FMI1(2)}_2\ {\delta }_{i,j\pm 2})]\hat{\alpha }^{{\dagger }}_i\hat{\alpha }_j
\end{equation}

\begin{equation} \label{GrindEQ__15_} 
\begin{array}{ll}
\hat{H}_{AFMI}=&\sum_{i,j}[{(A^{AFMI}_0\ {\delta }_{i,j}+A^{AFMI}_2\ {\delta }_{i,j\pm 2})\hat{\alpha }^{\dagger }_i\hat{\alpha }_j}\\
& +(B^{AFMI}_0{\delta }_{i,j}+B^{AFMI}_2{\delta }_{i,j\pm 2})\hat{\beta }^{\dagger }_i\hat{\beta }_j]
\end{array}
\end{equation}

\begin{equation}
\begin{array}{ll}
\hat{H}_{FMI1(2),AFMI}=&J_{FMI1(2),AFMI}(\hat{\alpha }^{\dagger }_{end(1),FMI1(2)}\\
&\hat{\alpha }_{1(end),AFMI}+\hat{\alpha }^{\dagger }_{{end(1),FMI1(2) }}\\
&\hat{\beta }_{1(end),AFMI})+H.c. 
\end{array}
\end{equation}
, and for antiparallel state, all the $\hat{\alpha}$ ($\hat{\alpha}^\dagger$) in Hamiltonian of $\hat{H}_{FMI2}$ and $\hat{\alpha}_{FMI2}$ ($\hat{\alpha}^\dagger_{FMI2}$) in Hamiltonian of $\hat{H}_{FMI2,AFMI}$ are replaced by $\hat{\beta}$ ($\hat{\beta}^\dagger$) and $\hat{\beta}_{FMI2}$ ($\hat{\beta}^\dagger_{FMI2}$).


 


We can use Eqs. (11${\sim}$16) to calculate magnon currents in three parts of magnon junction. The boundary conditions are set to be that magnon currents are continuous at interface and there are no magnon current injected from NM1 to FMI1. The simulation parameters are set to be on-site energy $A^{FMI{1}}_0=A^{FMI{2}}_0=A^{AFMI}_0=B^{AFMI}_0=0.5$ eV, nearest transition energy $A^{FMI{1}}_1=A^{FMI{2}}_1=-0.5$ eV, $A^{AFMI}_1=B^{AFMI}_1=-0.25$ eV, coupling energy of two types of magnons $J_{FMI1,AFMI}=J_{FMI2,AFMI}=1$ eV, spin chemical potential of two HMs layer ${\mu }_{NM1}={\mu }_{NM2}=0$, temperature $k_BT_{NM1}=0.026$ eV, $T_{FMI1}=0.9{\ }T_{NM1}$, $T_{AFMI}=0.8{\ }T_{NM1}$, $T_{FMI2}=0.7{\ }T_{NM1}$, $T_{NM2}=0.6{\ }T_{NM1}$, total site number $N_{FMI1}=N_{AFMI}=N_{FMI2}=\ 20$, coupling with two HMs layers ${\eta }^{L(R)}=8$ and Gilbert damping constant ${\alpha }_{FMI1}={\alpha }_{FMI2}=0.01$, ${\alpha }_{AFMI}=0.001$ (Details of different part's self-energy are in Supplemental Material \cite{sup}). 
Boundary condition is a nonlinear system of first order equations, and we can get a rough solution of ${\mu }_{FMI1}$, ${\mu }_{AFMI}$, ${\mu }_{FMI2}$.

For parallel magnetization, magnon potentials are ${\mu }_{FMI1}=20$ meV, ${\mu }_{AFMI}=5.3$ meV, ${\mu }_{FMI2}=18$ meV, and the magnon current at the interface of FMI2 and HM2 is $6.53\times 10^{-4}$ eV; for antiparallel magnetization state, magnon potentials are ${\mu }_{FMI1}=-37$ meV, ${\mu }_{AFMI}=5.2$ meV, ${\mu }_{FMI2}=-36.8$ meV, and the magnon current at the interface of FMI2 and HM2 is $4.79\times {10}^{-7}$ eV. It shows near 100 \% magnon junction ratio, here magnon junction ratio is MJR = $(J_{m,\uparrow \uparrow}-J_{m,\uparrow \downarrow})/(J_{m,\uparrow \uparrow}+J_{m,\uparrow \downarrow})$, where $J_{m,\uparrow \uparrow}$ and $J_{m,\uparrow \downarrow}$ are output magnon current of parallel magnetization and antiparallel magnetization state.

In conclusion, we propose a Green's function formalism to investigate magnon transport in AFMI or FIMI, which is a full quantum theory to study magnon transport in two-sublattice magnetic insulators. We studied the spatial distribution and temperature dependence of magnon current generated by the temperature or spin chemical potential step in FIMI or AFMI. Our results reveal that the magnon currents in both sublattices exhibit a positive correlation with temperature, and in AFMI, the magnon currents generated by temperature step in the two sublattices cancel each other out. Furthermore, we numerically simulate the magnon junction effect using the Green's function formalism, which yields a near 100 \% magnon junction ratio. Our work demonstrates the potential of employing full quantum theory to study magnon transport in specific magnonic devices.

This work is financial supported by the National Key Research and Development Program of China (MOST) (Grant No. 2022YFA1402800), the National Natural Science Foundation of China (NSFC) (Grant No. 51831012, 12134017), and partially supported by the Strategic Priority Research Program (B) (Grant No. XDB33000000).

\nocite{*}


\providecommand{\noopsort}[1]{}\providecommand{\singleletter}[1]{#1}%

\end{document}